\documentclass[draft,showkeys,showpacs,eqsecnum,nofootinbib,aps]{revtex4}
\renewcommand{\theequation}{\arabic{equation}}
\def\beq{\begin{equation}}
\def\eeq{\end{equation}}
\def\bea{\begin{eqnarray}}
\def\eea{\end{eqnarray}}\def\nn{\nonumber}

\def\na{\nabla}
\def\pa{\partial}

\def\nn{\nonumber}

\begin{document}
\title{Dynamics of stringy congruence in early universe}
\author{Yong Seung Cho}
\email{yescho@ewha.ac.kr} \affiliation{Department of Mathematics, Ewha Womans
University, Seoul 120-750 Korea}
\author{Soon-Tae Hong}
\email{soonhong@ewha.ac.kr} \affiliation{Department of Science
Education and Research Institute for Basic Sciences, Ewha Womans
University, Seoul 120-750 Korea}
\date{March 1, 2011}
\begin{abstract}
We studied the singularity of the geodesic surface congruence for timelike and null strings using the expansion of the universe in the string theory.  We had Raychaudhuri type equation for the expansion. Assuming the stringy strong energy condition and initial convergence, we induced the existence of singularity and got the same inequality equation of the string strong energy condition for both timelike and null stringy congruence. In this
paper we want to study the twist and shear aspects of the stingy geodesic surface congruence. Under some natural conditions we derive
the equations of the twist and the shear in terms of the expansion of the universe. In appendix we investigate the geodesic surface congruence of the null strings.
\end{abstract}
\maketitle

\section{Introduction}
\setcounter{equation}{0}
\renewcommand{\theequation}{\arabic{section}.\arabic{equation}}

The Hawking-Penrose (HP) singularity~\cite{hawking70} is assumed to exist at the
beginning of the universe. In the standard inflationary cosmology based on the HP
singularity theorem and inflationary scenario, the universe is
believed to expand from the Big Bang. Assuming that the early universe was filled
with a perfect fluid consisting of massive particles and/or massless particles and using
the strong energy condition which was used to show the HP singularity theorem,
one could find equations of state for each particle.

In the inflationary standard cosmology, it is believed that, after the Big Bang
explosion, radiation dominated phase occurred followed by matter dominated one,
even though there was a hot thermalization period of radiation and matter immediately
after the Big Bang. Moreover, a phase transition exists between massive particle
and massless particle phases in the universe. The equation of state of the massive
particle is different from that of the massless particle, and thus the massive particle
phase is not the same as the massless particle one.

Recently, applying the string theory~\cite{witten87,pol98} to cosmology, both of us
have studied the expansion of the universe in terms of the HP singularity in geodesic
surface congruences for the time-like and null strings~\cite{cho08,cho10}.
Taking an {\it ansatz} that the expansion of the stringy congruence is constant along the string
coordinate direction, we derived the Raychaudhuri type equation, which is an evolution
equation for the expansion, possessing correction terms associated with the stringy configurations.
Assuming the stringy strong energy condition, we have the HP type inequality equation which
produces the same inequality equation for both the time-like and null stringy congruences.

Nowadays, there have been some progresses in geometrical approaches to the
theoretical physics associated with the stringy congruence cosmology~\cite{cho08,cho10,cho10ijgm},
the stringy Jacobi-Morse theory~\cite{cho07}, the Sturm-Liouville theory~\cite{cho10jkps} and the
Gromov-Witten invariants~\cite{cho08jgm}. The variation of the surface spanned by closed strings
in a spacetime manifold was considered to discuss conjugate strings on the geodesic surface
and to induce the geodesic surface equation and the geodesic surface deviation equation, which
yields a Jacobi field and the index form of a geodesic surface as in the case of point particles~\cite{cho07}.
Later, after the geodesic equation and geodesic deviation equation with breaks on the path were formulated,
the physical changes of the action were investigated through the study of
the geometry of the moduli space associated with the critical points of the action functional and the
asymptotic boundary conditions in path space for point particles
in a conservative physical system, where the particle motion on the $n$-sphere $S^{n}$ was considered
to discuss the moduli space of the path space, the corresponding homology groups and the Sturm-Liouville
operators~\cite{cho10jkps}. Using symplectic cut-and-gluing formulae of the relative Gromov-Witten invariants,
one of us obtained a recursive formula for the Hurwitz number of
triple ramified geodesic surface coverings of a Riemann surface by
a Riemann surface~\cite{cho08jgm}.

In this paper, we extend the previous results in the stringy cosmology to study
the twist and shear aspects of the stringy geodesic congruences in the early universe.
To do this, we exploit the paradigm which can delineate the stringy
features of the HP singularity in the mathematical cosmology. Especially, we investigate
the effects of the twist and the shear of the stringy congruence on the ensuing universe evolution.

This paper is organized as follows. In Section II, we introduce the formalism
which describes the stringy congruence in the early universe.
In Section III, we briefly recapitulate the expansion rate of the timelike stringy congruence by
exploiting the Raychaudhuri type equation.
In Section IV, we investigate the aspects of twist and shear of the stringy
congruences in the early universe. Section V includes summary and discussions.
In Appendix A, we treat the null stringy congruence in the early universe.

\section{Congruence of strings}
\setcounter{equation}{0}
\renewcommand{\theequation}{\arabic{section}.\arabic{equation}}

The action for a string is proportional to the area of the surface
spanned in spacetime manifold $M$ by the evolution of the string.
In order to define the action on the curved manifold, we let $(M,
g_{ab})$ be a $D$-dimensional manifold associated with the metric
$g_{ab}$.  Given $g_{ab}$, we can have a unique covariant
derivative $\na_{a}$ satisfying~\cite{wald84} \bea
\na_{a}g_{bc}&=&0,\nn\\
\na_{a}\omega^{b}&=&\pa_{a}\omega^{b}+\Gamma^{b}_{~ac}~\omega^{c},\nn\\
(\na_{a}\na_{b}-\na_{b}\na_{a})\omega_{c}&=&R_{abc}^{~~~d}~\omega_{d}.\label{rtensor}
\eea
We parameterize the surface generated by the evolution of a string
by two world sheet coordinates $\tau$ and $\sigma$, and then we
have the corresponding vector fields $\xi^{a}=(\pa/\pa\tau)^{a}$
and $\zeta^{a}=(\pa/\pa\sigma)^{a}$. Since we have gauge degrees
of freedom, we can choose the orthonormal gauge as
follows~\cite{scherk75} \beq
\xi\cdot\zeta=0,~~~\xi\cdot\xi+\zeta\cdot\zeta=0,\label{gauge}\eeq
where the plus sign in the second equation is due to the fact that
$\xi\cdot\xi=-1$ is timelike and $\zeta\cdot\zeta=1$ is spacelike.
In the orthonormal gauge, we introduce tensor fields $B_{ab}$ and
$\bar{B}_{ab}$ defined as \beq B_{ab}=\na_{b}\xi_{a},~~~
\bar{B}_{ab}=\na_{b}\zeta_{a},\label{babbbabt}\eeq which satisfy the following
identities \bea B_{ab}\xi^{a}=0,~~~\bar{B}_{ab}\zeta^{a}&=&0,\nn\\
-B_{ab}\xi^{b}+\bar{B}_{ab}\zeta^{b}&=&0. \eea  Here in the last
equation, we have used the geodesic surface equation \beq
-\xi^{a}\na_{a}\xi^{b}+\zeta^{a}\na_{a}\zeta^{b}=0.\label{geodesic}\eeq
In particular, the timelike curves of the strings are geodesic,
then the geodesic surface equation holds.

We let the vector field $\eta^{a}=(\pa/\pa \alpha)^{a}$ be the
deviation vector which represents the displacement to an
infinitesimally nearby world sheet, and we let $\Sigma$ denote the
three-dimensional submanifold spanned by the world sheets
$\gamma_{\alpha}(\tau,\sigma)$. We then may choose $\tau$,
$\sigma$ and $\alpha$ as coordinates of $\Sigma$ to yield the
commutator relations, \bea
\pounds_{\xi}\eta^{a}&=&\xi^{b}\na_{b}\eta^{a}-\eta^{b}\na_{b}\xi^{a}=0,\nn\\
\pounds_{\zeta}\eta^{a}&=&\zeta^{b}\na_{b}\eta^{a}-\eta^{b}\na_{b}\zeta^{a}=0,\nn\\
\pounds_{\xi}\zeta^{a}&=&\xi^{b}\na_{b}\zeta^{a}-\zeta^{b}\na_{b}\xi^{a}=0.\label{poundxizeta}
\eea Using the above relations, we obtain \beq
\xi^{a}\na_{a}\eta^{b}-\zeta^{a}\na_{a}\eta^{b}=(B^{b}_{~a}-\bar{B}^{b}_{~a})\eta^{a}.\eeq

Next we introduce the metrics $h_{ab}$ and $\bar{h}_{ab}$, \beq
h_{ab}=g_{ab}+\xi_{a}\xi_{b},~~~\bar{h}_{ab}=g_{ab}-\zeta_{a}\zeta_{b},\label{projections}\eeq
which satisfy \beq
\begin{array}{lll}
h_{ab}\xi^{a}=0, &h_{ab}\xi^{b}=0,
&h_{ab}g^{bc}h_{cd}=h_{ad},\\
\bar{h}_{ab}\zeta^{a}=0, &\bar{h}_{ab}\zeta^{b}=0,
&\bar{h}_{ab}g^{bc}\bar{h}_{cd}=\bar{h}_{ad},\\
h_{ab}h^{ab}=D-1, &\bar{h}_{ab}\bar{h}^{ab}=D-1,
&h_{ab}\bar{h}^{ab}=D-2.
\end{array} \eeq
Here note that $h_{ab}$ and $\bar{h}_{ab}$ are the metrics on the
hypersurfaces orthogonal to $\xi^{a}$ and $\zeta^{a}$,
respectively. Moreover, we can define projection operators
$h^{a}_{~b}$ and $\bar{h}^{a}_{~b}$ as follows \beq
h^{a}_{~b}=g^{ac}h_{cb},~~~\bar{h}^{a}_{~b}=g^{ac}\bar{h}_{cb}.\eeq
These operators fulfil \beq \begin{array}{cc}
h^{a}_{~b}h^{b}_{~c}=h^{ab}h_{bc}=h^{a}_{~c},
&\bar{h}^{a}_{~b}\bar{h}^{b}_{~c}=\bar{h}^{ab}\bar{h}_{bc}=\bar{h}^{a}_{~c},\\
h_{ab}h^{bc}h_{cd}=h_{ad},
&\bar{h}_{ab}\bar{h}^{bc}\bar{h}_{cd}=\bar{h}_{ad}.
\end{array}\eeq

Now, we decompose $B_{ab}$ into three pieces \beq
B_{ab}=\frac{1}{D-1}\theta h_{ab}+\sigma_{ab}+\omega_{ab},\label{bab}\eeq
where the expansion $\theta$, the shear $\sigma_{ab}$ and the twist $\omega_{ab}$ of the
stringy congruence are
given by
\beq\theta=B^{ab}h_{ab},~~~
\sigma_{ab}=B_{(ab)}-\frac{1}{D-1}\theta h_{ab},~~~
\omega_{ab}=B_{[ab]}.\eeq Similarly, $\bar{B}_{ab}$ is also
decomposed into three parts \beq
\bar{B}_{ab}=\frac{1}{D-1}\bar{\theta}
\bar{h}_{ab}+\bar{\sigma}_{ab}+\bar{\omega}_{ab},\eeq where
\beq\bar{\theta}=\bar{B}^{ab}\bar{h}_{ab},~~~
\bar{\sigma}_{ab}=\bar{B}_{(ab)}-\frac{1}{D-1}\bar{\theta}
\bar{h}_{ab},~~~\bar{\omega}_{ab}=\bar{B}_{[ab]}.\label{barbarbar}\eeq We then find
\beq
\begin{array}{cc}
\sigma_{ab}h^{ab}=0, &\omega_{ab}h^{ab}=0,\\
\bar{\sigma}_{ab}\bar{h}^{ab}=0, &\bar{\omega}_{ab}\bar{h}^{ab}=0,\\
-\sigma_{ab}\xi^{b}+\bar{\sigma}_{ab}\zeta^{b}=0,
&-\omega_{ab}\xi^{b}+\bar{\omega}_{ab}\zeta^{b}=0,
\end{array} \eeq and
\beq -\xi^{c}\na_{c}B_{ab}+\zeta^{c}\na_{c}\bar{B}_{ab}
=B^{c}_{~b}B_{ac}-\bar{B}^{c}_{~b}\bar{B}_{ac}-R_{cbad}(\xi^{c}\xi^{d}-\zeta^{c}\zeta^{d}).
\label{bbr}\eeq

Exploiting (\ref{bbr}) one arrives at \bea
-\xi^{a}\na_{a}\theta+\zeta^{a}\na_{a}\bar{\theta}
&=&\frac{1}{D-1}(\theta^{2}-\bar{\theta}^{2})+\sigma_{ab}\sigma^{ab}-\bar{\sigma}_{ab}\bar{\sigma}^{ab}
-\omega_{ab}\omega^{ab}+\bar{\omega}_{ab}\bar{\omega}^{ab}+R_{ab}(\xi^{a}\xi^{b}-\zeta^{a}\zeta^{b}),\label{eq1}\\
-\xi^{c}\na_{c}\omega_{ab}+\zeta^{c}\na_{a}\bar{\omega}_{ab}
&=&\frac{2}{D-1}\theta(\omega_{ab}-\xi^{c}\xi_{[a}\omega_{b]c})
-\frac{2}{D-1}\bar{\theta}(\bar{\omega}_{ab}+\zeta^{c}\zeta_{[a}\bar{\omega}_{b]c})
+2(\sigma^{c}_{~[b}\omega_{a]c}-\bar{\sigma}^{c}_{~[b}\bar{\omega}_{a]c}),\label{eq2}\\
-\xi^{c}\na_{c}\sigma_{ab}+\zeta^{c}\na_{a}\bar{\sigma}_{ab}
&=&\frac{1}{(D-1)^{2}}(\theta^{2}\xi_{a}\xi_{b}+\bar{\theta}^{2}\zeta_{a}\zeta_{b})
+\frac{2}{D-1}(\theta h^{c}_{~(a}-\bar{\theta} \bar{h}^{c}_{~(a})\sigma_{b)c}+\sigma_{ac}\sigma^{c}_{~b}
-\bar{\sigma}_{ac}\bar{\sigma}^{c}_{~b}\nn\\
&&+\omega_{ac}\omega^{c}_{~b}-\bar{\omega}_{ac}\bar{\omega}^{c}_{~b}
-\left(R_{c(ab)d}+\frac{1}{D-1}g_{ab}R_{cd}\right)(\xi^{a}\xi^{b}-\zeta^{a}\zeta^{b})\nn\\
&&-\frac{1}{D-1}g_{ab}(\sigma_{cd}\sigma^{cd}-\bar{\sigma}_{cd}\bar{\sigma}^{cd}
-\omega_{cd}\omega^{cd}+\bar{\omega}_{cd}\bar{\omega}^{cd})
+\frac{1}{D-1}\theta\xi^{c}\xi_{(a}\na_{|c|}\xi_{b)}\nn\\
&&+\frac{1}{D-1}\bar{\theta}\zeta^{c}\zeta_{(a}\na_{|c|}\zeta_{b)}
+\frac{1}{D-1}\xi_{a}\xi_{b}\xi^{c}\na_{c}\theta
+\frac{1}{D-1}\zeta_{a}\zeta_{b}\zeta^{c}\na_{c}\bar{\theta}.
\label{eq3}
\eea

\section{Expansion of stringy congruence}
\setcounter{equation}{0}
\renewcommand{\theequation}{\arabic{section}.\arabic{equation}}

The motion types of stringy congruence can be described in terms of expansion, twist and shear.
In this section, we will pedagogically summarize the previous results~\cite{cho08,cho10} on the expansion rate of
stringy congruence in the early universe for the sake of completeness.  We will consider the twist and shear
motions in the next section.

Taking an {\it ansatz} that the expansion $\bar{\theta}$ is constant
along the $\sigma$-direction, from (\ref{eq3}) one obtains a
Raychaudhuri type equation \beq
\frac{d\theta}{d\tau}=-\frac{1}{D-1}(\theta^{2}-\bar{\theta}^{2})
-\sigma_{ab}\sigma^{ab}+\bar{\sigma}_{ab}\bar{\sigma}^{ab}
+\omega_{ab}\omega^{ab} -\bar{\omega}_{ab}\bar{\omega}^{ab}
-R_{ab}(\xi^{a}\xi^{b}-\zeta^{a}\zeta^{b}).\label{treq} \eeq We
now assume that $\omega_{ab}=\bar{\omega}_{ab}$,
$\sigma_{ab}=\bar{\sigma}_{ab}$ and a
stringy strong energy condition \beq
R_{ab}(\xi^{a}\xi^{b}-\zeta^{a}\zeta^{b})=8\pi\left(T_{ab}
(\xi^{a}\xi^{b}-\zeta^{a}\zeta^{b})+\frac{2}{D-2}T\right)\ge 0,
\label{strong}\eeq where $T_{ab}$ and $T$ are the energy-momentum
tensor and its trace, respectively. The Raychaudhuri type equation
(\ref{treq}) then has a solution of the form \beq
\frac{1}{\theta(\tau)}\ge
\frac{1}{\theta(0)}+\frac{1}{D-1}\left(\tau-\int_{0}^{\tau} {\rm
d}\tau~\left(\frac{\bar{\theta}}{\theta}\right)^{2}\right).\label{soln}
\eeq We assume that $\theta(0)$ is negative so that the
congruence is initially converging as in the point particle case.
The inequality (\ref{soln}) implies that $\theta(\tau)$ must pass
through the singularity within a proper time \beq \tau\le
\frac{D-1}{|\theta(0)|}+\int_{0}^{\tau} {\rm
d}\tau~\left(\frac{\bar{\theta}}{\theta}\right)^{2}.\label{propertime}
\eeq For a perfect fluid, the energy-momentum tensor given by
\beq
T_{ab}=\rho~u_{a}u_{a}+P~(g_{ab}+u_{a}u_{b})
\eeq
where $\rho$ and
$P$ are the mass-energy density and pressure of the fluid as
measured in its rest frame, respectively, and $u^{a}$ is the
time-like $D$-velocity in its rest frame~\cite{mtw,wald84}, the
stringy strong energy condition (\ref{strong}) yields only one
inequality equation \beq \frac{D-4}{D-2}\rho+\frac{D}{D-2}P\ge 0.
\label{secstring} \eeq

Now, we consider the point particle limit of the timelike stringy congruence.
If the fiber space $F$ in the fibration $\pi:
M\rightarrow N_{4}$ is a point, then the total space $M$ is the same
as the base spacetime four manifold $N_{4}$.  In this case, the
geodesic surfaces are geodesic in $N_{4}$, the congruence of time-like
geodesic surfaces is a congruence of time-like geodesics, and so
$\bar{B}_{ab}=\bar{\theta}=\bar{\sigma}_{ab}=\bar{\omega}_{ab}=0$.
If the congruence is hypersurface orthogonal, then we have
$\omega_{ab}=0$.  Suppose that the strong energy condition
$R_{ab}\xi^{a}\xi^{b}\ge 0$ is satisfied to yield two
inequalities~\cite{hawking70,wald84,carroll} \beq
\rho+3P\ge0,~~~\rho+P\ge0.\label{strongpoint}\eeq We then have the
differential inequality equation
\beq
\frac{d\theta}{d\tau}+\frac{1}{3}\theta^{2}\le 0
\eeq which has a
solution in the following form
\beq\frac{1}{\theta(\tau)}\ge
\frac{1}{\theta(0)}+\frac{1}{3}\tau.\eeq If we assume that
$\theta(0)$ is negative, the expansion $\theta(\tau)$ must go to the
negative infinity along that geodesic within a proper time
\beq\tau\le \frac{3}{|\theta(0)|},\eeq  whose consequence coincides with
that of Hawking and Penrose~\cite{hawking70}.

Next, we consider the expansion of the null stringy congruence in the early universe, which is
explicitly described in Appendix A. Taking the {\it ansatz} that the expansion $\bar{\theta}$ is constant
along the $\sigma$-direction as in the time-like case, we have
another Raychaudhuri type equation (\ref{treqnull}). With the assumption that
$\omega_{ab}=\bar{\omega}_{ab}$,
$\sigma_{ab}=\bar{\sigma}_{ab}$ and
a stringy strong energy condition (\ref{rkknull}) for null case,
exploiting the energy-momentum tensor of the perfect fluid
we reproduce the inequality (\ref{secstring}) in the time-like congruence of
strings. We assume again that $\theta(0)$ is negative. The
inequality (\ref{solnnull}) then implies that $\theta(\tau)$ must pass
through the singularity within an affine length
$$\lambda\le
\frac{D-2}{|\theta(0)|}+\frac{D-2}{D-1}\int_{0}^{\lambda} {\rm
d}\lambda~\left(\frac{\bar{\theta}}{\theta}\right)^{2}
$$
as in (\ref{propertimenull}).

In the point particle limit with the
strong energy condition
$$R_{ab}k^{a}k^{b}\ge 0$$
in (\ref{rkknull}), one can obtain the equation of state
$$\rho+P\ge 0$$
in (\ref{strongpointnull}) for the null point congruence~\cite{hawking70,wald84,carroll}. If we assume
that the initial value is negative, the expansion $\theta(\tau)$ must go
to the negative infinity along that geodesic within a finite
affine length
$$
\lambda\le\frac{2}{|\theta(0)|}
$$
as in (\ref{lambdanullpoint})~\cite{hawking70}.

Moreover, the stringy universe evolves without any phase transition, since there exists
only one equation of state (\ref{secstring}) both for the radiation and matter, differently from the
point particle inflationary cosmology with two equations of state in (\ref{strongpoint}) and
(\ref{strongpointnull}) for matter and radiation, respectively. 


\section{Twist and shear of stringy congruence}
\setcounter{equation}{0}
\renewcommand{\theequation}{\arabic{section}.\arabic{equation}}

In this section we will consider the twist and shear of stringy
congruence in the early universe. First, we investigate the twist
feature of the stringy congruence. Taking in (\ref{eq3}) an {\it ansatz}
that the twist $\bar{\omega}_{ab}$ is constant along the
$\sigma$-direction, we obtain an evolution equation for the twist
\beq
\frac{d\omega_{ab}}{d\tau}=-\frac{2}{D-1}\theta(\omega_{ab}-\xi^{c}\xi_{[a}\omega_{b]c})
+\frac{2}{D-1}\bar{\theta}(\bar{\omega}_{ab}+\zeta^{c}\zeta_{[a}\bar{\omega}_{b]c})
-2(\sigma^{c}_{~[b}\omega_{a]c}-\bar{\sigma}^{c}_{~[b}\bar{\omega}_{a]c}).
\eeq

We now assume that $\omega_{ab}=\bar{\omega}_{ab}$,
$\sigma_{ab}=\bar{\sigma}_{ab}$ and $\theta\gg\bar{\theta}$ to
obtain\footnote{In deriving (\ref{soln}), we did not neglect the
$\bar{\theta}$ correction terms. However, from now on, we will keep
the zeroth order term of $\bar{\theta}$ with respect to $\theta$
to see the twist and shear features of the stringy congruence.}
\beq \frac{d\omega_{ab}}{d\tau}=-\frac{2}{D-1}\theta(\omega_{ab}
-\xi^{c}\xi_{[a}\omega_{b]c}).\label{omegatau} \eeq
Here one notes that the twist $\omega_{bc}$ is orthogonal to the timelike vector
field $\xi^{c}$ so that their inner product contraction $\xi^{c}\omega_{bc}$
in (\ref{omegatau}) vanishes. We then have the above equation of the form
\beq \frac{d\omega_{ab}}{d\tau}=-\frac{2}{D-1}\theta\omega_{ab}\label{omegatau2} \eeq
which has solution of the form
\beq
\omega_{ab}(\tau)=\omega_{ab}(0)\exp\left(-\frac{2}{D-1}\int_{0}^{\tau}{\rm d}\tau~\theta\right).
\label{omehaabtime}
\eeq
This solution indicates that, as the early universe evolves with the expansion rate $\theta$,
$\theta$ increases and the twist of the stringy congruence $\omega_{ab}$ decreases exponentially.
Moreover, the initial twist $\omega_{ab}(0)$ should be enormously large enough to support the whole
rotations of the ensuing universe later.

It is worthy to note that in the higher $D$-dimensional stringy cosmology, one can
have the condition $\omega_{ab}=\bar{\omega}_{ab}\neq 0$, where the nonvanishing $\omega_{ab}$ initiates
the rotational degrees of freedom in the universe such as the rotational motions of galaxies, stars, planets and moons.
Moreover the nonvanishing $\bar{\omega}_{ab}$ could explain the rotational
degrees of freedom of the strings or physical particles themselves~\cite{scherk75,witten87,pol98}.
Next, we consider the point particle limit of the timelike stringy congruence where
$\omega_{ab}=\bar{\omega}_{ab}=0$. We can then have the Hawking and
Penrose limit with $\omega_{ab}=0$ in the $D=4$ point particle
congruence cosmology~\cite{hawking70}.

Second, we study the shear of the stringy congruence. Taking an
{\it ansatz} that the shear $\bar{\sigma}_{ab}$ is constant along the
$\sigma$-direction, from (\ref{eq3}) we obtain an evolution
equation for the shear. \bea
\frac{d\sigma_{ab}}{d\tau}
&=&-\frac{1}{(D-1)^{2}}(\theta^{2}\xi_{a}\xi_{b}+\bar{\theta}^{2}\zeta_{a}\zeta_{b})
-\frac{2}{D-1}(\theta h^{c}_{~(a}-\bar{\theta} \bar{h}^{c}_{~(a})\sigma_{b)c}-\sigma_{ac}\sigma^{c}_{~b}
+\bar{\sigma}_{ac}\bar{\sigma}^{c}_{~b}\nn\\
&&-\omega_{ac}\omega^{c}_{~b}+\bar{\omega}_{ac}\bar{\omega}^{c}_{~b}
+\left(R_{c(ab)d}+\frac{1}{D-1}g_{ab}R_{cd}\right)(\xi^{a}\xi^{b}-\zeta^{a}\zeta^{b})\nn\\
&&+\frac{1}{D-1}g_{ab}(\sigma_{cd}\sigma^{cd}-\bar{\sigma}_{cd}\bar{\sigma}^{cd}
-\omega_{cd}\omega^{cd}+\bar{\omega}_{cd}\bar{\omega}^{cd})
-\frac{1}{D-1}\theta\xi^{c}\xi_{(a}\na_{|c|}\xi_{b)}\nn\\
&&-\frac{1}{D-1}\bar{\theta}\zeta^{c}\zeta_{(a}\na_{|c|}\zeta_{b)}
-\frac{1}{D-1}\xi_{a}\xi_{b}\xi^{c}\na_{c}\theta
-\frac{1}{D-1}\zeta_{a}\zeta_{b}\zeta^{c}\na_{c}\bar{\theta}.
\label{eq33}
\eea
We again assume that $\omega_{ab}=\bar{\omega}_{ab}$,
$\sigma_{ab}=\bar{\sigma}_{ab}$ and $\theta\gg\bar{\theta}$ to
yield \bea
\frac{d\sigma_{ab}}{d\tau}
&=&-\frac{1}{(D-1)^{2}}\theta^{2}\xi_{a}\xi_{b}
-\frac{2}{D-1}\theta h^{c}_{~(a}\sigma_{b)c}
+\left(R_{c(ab)d}+\frac{1}{D-1}g_{ab}R_{cd}\right)(\xi^{a}\xi^{b}-\zeta^{a}\zeta^{b})\nn\\
&&-\frac{1}{D-1}\theta\xi^{c}\xi_{(a}\na_{|c|}\xi_{b)}-\frac{1}{D-1}\xi_{a}\xi_{b}\xi^{c}\na_{c}\theta.
\label{sigmaabdyn}
\eea

At this point, we digress to carefully consider the shear tensor field $\sigma_{ab}$ of the stringy congruence.
In the $D$-dimensional spacetime manifold $(M, g_{ab})$, we considered the metrics $h_{ab}$ and $\bar{h}_{ab}$ in
(\ref{projections}) on the hypersurfaces orthogonal to the timelike direction and to the string direction, respectively.
The metrics $g_{ab}$, $h_{ab}$ and $\bar{h}_{ab}$ have signatures $(1, D-1)$, $(0, D-1)$ and $(1, D-2)$, respectively.
In particular, $h_{ab}$ is positive definite and may have an Euclidean metric on the $(D-1)$-dimensional hypersurface $N_{D-1}$
which is orthogonal to the time direction. We may now choose orthogonal basis for the hypersurface $N_{D-1}$.

The symmetric part $B_{(ab)}$ of the tensor field $B_{ab}$ on the hypersurface $N_{D-1}$ is given by a $(D-1)\times(D-1)$ matrix
which can be split into two pieces as follows
\beq
B_{(ab)}=\frac{1}{D-1}\theta h_{ab}+\sigma_{ab},
\eeq
where
\beq
\frac{1}{D-1}\theta h_{ab}=\left(
\begin{array}{ccc}
\frac{\theta}{D-1} & &\\
 &\cdots &\\
  & &\frac{\theta}{D-1}
\end{array}
\right),~~~
\sigma_{ab}=\left(
\begin{array}{ccc}
\theta_{1}-\frac{\theta}{D-1} & &\sigma_{ij}\\
 &\cdots &\\
\sigma_{ji} & &\theta_{D-1}-\frac{\theta}{D-1}
\end{array}
\right).
\eeq
Here $\sigma_{ij}$ are off-diagonal elements of the matrix $\sigma_{ab}$.
It is well known in astrophysics that the universe is homogeneously and isotropically expanding.
Exploiting the fact that the universe is homogeneously expanding, one can see that the off-diagonal part of the shear
tensor vanishes, $\sigma_{ij}=0$, to yield
\beq
\sigma_{ab}={\rm diag}~\left(\theta_{1}-\frac{\theta}{D-1},\cdots,\theta_{D-1}-\frac{\theta}{D-1}\right)
\eeq
Next, since the universe is isotropically expanding, one can observe that all the diagonal elements of the shear
tensor are the same so that we can arrive at
\beq
\theta_{1}=\cdots=\theta_{D-1}.
\eeq
Moreover, by definition the shear tensor field $\sigma_{ab}$ is traceless and symmetric to yield
\beq
\theta_{a}=\frac{\theta}{D-1},~~~(a=1,2,\cdots,D-1)
\eeq
which indicates that all the shear tensor components vanishes,
\beq
\sigma_{ab}=0.
\eeq
This result on Euclidean manifold can be extended to the more general curved manifold case without loss of generality.
One can thus conclude that there are no shear features in the homogeneous and isotropic universe
regardless of the dynamic equation for the shear $\sigma_{ab}$ in (\ref{sigmaabdyn}). Next, we consider
the point particle limit of the timelike stringy congruence in which $\sigma_{ab}=\bar{\sigma}_{ab}=0$.
In this case we can have the Hawking and Penrose limit with $\sigma_{ab}=0$ in the point particle
congruence cosmology~\cite{hawking70}.

Now, we consider the twist and shear of the null stringy congruence in the early universe, which is systematically
delineated in Appendix A. Exploiting the fact that the twist $\omega_{bc}$ is orthogonal to the null tangent vector field
$k^{c}$, one can arrive at the evolution equation
$$
\omega_{ab}(\lambda)=\omega_{ab}(0)\exp\left(-\frac{2}{D-2}\int_{0}^{\lambda}{\rm d}\lambda~\theta\right)
$$
as in (\ref{omegaabnull}) of the twist $\omega_{ab}$ of the null stringy congruence along the affine parameter $\lambda$.
This shows that $\omega_{ab}$ decreases exponentially with the modified factor associated with the dimensionality,
with respect to the timelike stringy congruence in (\ref{omehaabtime}).

Next, in order to consider the point particle limit of the null stringy congruence,
we first assume that $\omega_{ab}=\bar{\omega}_{ab}$,
$\sigma_{ab}=\bar{\sigma}_{ab}$ and $\theta\gg\bar{\theta}$ to
yield (\ref{sigmaabdynnull}). As in the case of the time-like stringy congruence,
all the shear tensor components again vanishes as in (\ref{sigmaabnull}), so that
there are no shear features in the homogeneous and isotropic universe
regardless of the dynamic equation for the shear $\sigma_{ab}$ in (\ref{sigmaabdynnull}).
As for the point particle case of the twist of the null congruence, we have
$\omega_{ab}=\bar{\omega}_{ab}=\sigma_{ab}=\bar{\sigma}_{ab}=0$
so that there are no twist and shear motions in the homogeneous and isotropic universe.

\section{Conclusions}
\setcounter{equation}{0}
\renewcommand{\theequation}{\arabic{section}.\arabic{equation}}

In summary, the stringy universe evolves without any phase transition, because there is
only one equation of state both for the radiation and matter, differently from the
point particle inflationary cosmology with two equations of state for matter and radiation,
respectively. By exploiting the fact that there is only one equation of state in evolution
of the universe, it was also shown that the stringy cosmology is cyclic, similar to the
brane cyclic cosmology, but modified: Big Bang, radiation-matter mixture phase, dark
energy dominated phase, Big Crunch, and again Big Bang.

In the higher dimensional stringy cosmology, as the early universe
evolves with the expansion rate $\theta$, $\theta$ increases and
the twist of the stringy congruence $\omega_{ab}$ decreases
exponentially, and the initial twist $\omega_{ab}(0)$ should be
extremely large enough to support the whole rotations of the
ensuing universe. It is worthy to note that in the stringy
cosmology one can have the condition
$\omega_{ab}=\bar{\omega}_{ab}\neq 0$. Here the nonvanishing
$\omega_{ab}$ initiates the rotational degrees of freedom in the
universe such as the rotational motions of galaxies, stars,
planets and moons, while the nonvanishing $\bar{\omega}_{ab}$
could explain the rotational degrees of freedom of the strings or
physical particles themselves. On the other hand, the effects of
the shear of the stringy congruence on the ensuing universe
evolution are negligible to produce the isotropic and homogeneous
universe features, regardless of the details of the dynamic
equations of motions for the shear of the stringy congruence.

Next, for the null stringy congruence corresponding to the
massless photons in the higher dimensional cosmology, through the
evolution of the early universe, the expansion rate $\theta$
increases and the twist $\omega_{ab}$ of the null stringy
congruence $\omega_{ab}$ decreases exponentially, and the initial
twist is extremely large enough to generate the whole rotations of
the ensuing universe, similar to the case of the timelike stringy
congruence corresponding to the massive physical particles. In the
null stringy cosmology one can also have the condition
$\omega_{ab}=\bar{\omega}_{ab}\neq 0$. $\omega_{ab}$ initiates the
rotational degrees of freedom in the universe such as the
celestial body rotational motions, while the nonvanishing
$\bar{\omega}_{ab}$ could explain the rotational degrees of
freedom of the strings or physical photons themselves. On the
other hand, there exist no effects of the shear of the null
stringy congruence on the ensuing universe evolution to produce
the isotropic and homogeneous universe features, regardless of the
details of the dynamic equations of motions for the shear of the
stringy congruence.

Recently, the Alice detector of the Large Hadron Collider (LHC) is scheduled to
detect the so-called quark-gluon plasma state, which is assumed to
exist in an extremely hot soup of massive quarks and massless gluons. Both in the
standard and stringy cosmologies, this quark-gluon plasma state is
supposed to occur immediately after the Big Bang of the tiny early
universe manufactured in the LHC. In the point-particle standard cosmology,
the quark-gluon plasma state can exist shortly and disappear eventually to enter the radiation dominated phase,
while in the stringy higher dimensional cosmology the quark-gluon plasma state can
develop into particles such as protons and neutrons and sustain
the radiation and matter mixture phase. It is expected that the Alice
will be able to detect the procedure of particle states along with
the evolution of the tiny universe planned to occur at the LHC and
it will be able to determine which cosmology is viable.
We recall that as far as radiation and matter are concerned, the mixture of these two
exists together in the current universe.

\acknowledgments
The research of YSC was supported by Basic Science Research Program
through the National Research Foundation of Korea (NRF) funded by the Ministry
of Education, Science and Technology (2010-0011145).

\appendix
\section{Null stringy congruence}
\setcounter{equation}{0}
\renewcommand{\theequation}{A.\arabic{equation}}

In this section, we will investigate the congruence of the null strings, where the
tangent vector of a null curve is normal to itself. See
Refs.~\cite{schild77,karlhede86,roshchupkin95} for the proper
definition and propagation of the classical null strings. We
consider the evolution of vectors in a $(D-2)$-dimensional
subspace of spatial vectors normal to the null tangent vector
field $k^{a}=(\pa/\pa\lambda)^{a}$, where $\lambda$ is the affine
parameter, and to an auxiliary null vector $l^{a}$ which points in
the opposite spatial direction to $k^{a}$, normalized by~\cite{carroll}
\beq
l^{a}k_{a}=-1
\eeq and is parallel transported,
namely,
\beq
k^{a}\na_{a}l^{b}=0.
\eeq
The spatial vectors in the
$(D-2)$-dimensional subspace are then orthogonal to both $k^{a}$
and $l^{a}$.

We now introduce the metrics $n_{ab}$ and $\bar{h}_{ab}$ which is defined in (\ref{projections}), \beq
n_{ab}=g_{ab}+k_{a}l_{b}+l_{a}k_{b},~~~\bar{h}_{ab}=g_{ab}-\zeta_{a}\zeta_{b}.
\label{projection2}\eeq
Similarly to the time-like case, we introduce tensor fields
\beq
B_{ab}=\na_{b}k_{a},~~~\bar{B}_{ab}=\na_{b}\zeta_{a},
\eeq
satisfying the identities
\beq
B_{ab}k^{a}=\bar{B}_{ab}\zeta^{a}=0,~~~-B_{ab}k^{b}+\bar{B}_{ab}\zeta^{b}=0.
\eeq
We also define the
deviation vector $\eta^{a}=(\pa/\pa \alpha)^{a}$ representing the
displacement to an infinitesimally nearby world sheet so that we
can choose $\lambda$, $\sigma$, and $\alpha$ as coordinates of the
three-dimensional submanifold spanned by the world sheets. We then
have the commutator relations
\beq
\pounds_{k}\eta^{a}=\pounds_{\zeta}\eta^{a}=\pounds_{k}\zeta^{a}=0,~~~
k^{a}\na_{a}\eta^{b}-\zeta^{a}\na_{a}\eta^{b}=(B^{b}_{~a}-\bar{B}^{b}_{~a})\eta^{a}.
\eeq

We decompose $B_{ab}$ into three pieces \beq
B_{ab}=\frac{1}{D-2}\theta
n_{ab}+\sigma_{ab}+\omega_{ab},\label{babnull2}\eeq where the
expansion, shear, and twist of the stringy congruence along the
affine direction are defined as
\beq
\theta=B^{ab}n_{ab},~~~
\sigma_{ab}=B_{(ab)}-\frac{1}{D-2}\theta n_{ab},~~~
\omega_{ab}=B_{[ab]}.
\eeq
It is noteworthy that even though we have
the same notations for $B_{ab}$, $\theta$, $\sigma_{ab}$ and
$\omega_{ab}$ in (\ref{bab}) and (\ref{babnull2}), the differences
of these notations among the time-like sting cases and null string
cases are understood in the context. The metric $n_{ab}$ also satisfies the identities
\beq
\sigma_{ab}n^{ab}=\omega_{ab}n^{ab}=0,
\eeq
and
\beq
n_{ab}k^{a}=n_{ab}k^{b}=n_{ab}l^{a}=n_{ab}l^{b}=0,~~~
n_{ab}g^{bc}n_{cd}=n_{ad},~~~
n_{ab}n^{ab}=D-2,~~~
n_{ab}\bar{h}^{ab}=D-3.
\eeq
We define $n^{a}_{~b}$ as
\beq
n^{a}_{~b}=g^{ac}n_{cb}=\delta^{a}_{~b}+k^{a}l_{b}+l^{a}k_{b},
\eeq
which fulfills the following identities
\beq
k^{c}\na_{c}n^{a}_{~b}=0.
\eeq
and
\beq
n^{a}_{~b}k^{b}=n^{a}_{~b}k_{a}=n^{a}_{~b}l^{b}=n^{a}_{~b}l_{a}=0,~~~n^{a}_{~b}n^{b}_{~c}=n^{a}_{~c},~~~
n_{ab}n^{ac}=n_{b}^{~c},~~~n_{a}^{~b}n_{bc}=n_{ac}.
\eeq

Similarly, we decompose $\bar{B}_{ab}$ into three parts as in the time-like case
\beq
\bar{B}_{ab}=\frac{1}{D-1}\bar{\theta}
\bar{h}_{ab}+\bar{\sigma}_{ab}+\bar{\omega}_{ab},\eeq
where $\bar{\theta}$, $\bar{\sigma}_{ab}$ and $\bar{\omega}_{ab}$ are given by
(\ref{barbarbar}).  We then have the identities
\beq
B_{ab}k^{a}=\bar{B}_{ab}\zeta^{a}=0,~~~-\sigma_{ab}k^{b}+\bar{\sigma}_{ab}\zeta^{b}=0,~~~
-\omega_{ab}k^{b}+\bar{\omega}_{ab}\zeta^{b}=0,
\eeq
and
\beq
-k^{c}\na_{c}B_{ab}+\zeta^{c}\na_{c}\bar{B}_{ab}
=B^{c}_{~b}B_{ac}-\bar{B}^{c}_{~b}\bar{B}_{ac}-R_{cbad}(k^{c}k^{d}-\zeta^{c}\zeta^{d}).
\label{bbrnull}\eeq

Using (\ref{bbrnull}) we find
\bea
-k^{a}\na_{a}\theta+\zeta^{a}\na_{a}\bar{\theta}
&=&\frac{1}{D-2}\theta^{2}-\frac{1}{D-1}\bar{\theta}^{2}+\sigma_{ab}\sigma^{ab}-\bar{\sigma}_{ab}\bar{\sigma}^{ab}
-\omega_{ab}\omega^{ab}+\bar{\omega}_{ab}\bar{\omega}^{ab}+R_{ab}(k^{a}k^{b}-\zeta^{a}\zeta^{b}),\label{eq1null}\\
-k^{c}\na_{c}\omega_{ab}+\zeta^{c}\na_{a}\bar{\omega}_{ab}
&=&\frac{2}{D-2}\theta(\omega_{ab}-k^{c}k_{[a}\omega_{b]c})
-\frac{2}{D-1}\bar{\theta}(\bar{\omega}_{ab}+\zeta^{c}\zeta_{[a}\bar{\omega}_{b]c})
+2(\sigma^{c}_{~[b}\omega_{a]c}-\bar{\sigma}^{c}_{~[b}\bar{\omega}_{a]c}),\label{eq2null}\\
-k^{c}\na_{c}\sigma_{ab}+\zeta^{c}\na_{a}\bar{\sigma}_{ab}
&=&\frac{1}{(D-2)^{2}}\theta^{2}k_{a}k_{b}+\frac{1}{(D-1)^{2}}\bar{\theta}^{2}\zeta_{a}\zeta_{b}
+\frac{2}{D-2}\theta h^{c}_{~(a}\sigma_{b)c}
-\frac{2}{D-1}\bar{\theta} \bar{h}^{c}_{~(a}\sigma_{b)c}\nn\\
&&+\sigma_{ac}\sigma^{c}_{~b}
-\bar{\sigma}_{ac}\bar{\sigma}^{c}_{~b}+\omega_{ac}\omega^{c}_{~b}-\bar{\omega}_{ac}\bar{\omega}^{c}_{~b}
-\left(R_{c(ab)d}+\frac{1}{D-2}g_{ab}R_{cd}\right)k^{c}k^{d}\nn\\
&&+\left(R_{c(ab)d}+\frac{1}{D-1}g_{ab}R_{cd}\right)\zeta^{c}\zeta^{d}
-\frac{1}{D-2}g_{ab}(\sigma_{cd}\sigma^{cd}-\omega_{cd}\omega^{cd})\nn\\
&&+\frac{1}{D-1}g_{ab}(\bar{\sigma}_{cd}\bar{\sigma}^{cd}
-\bar{\omega}_{cd}\bar{\omega}^{cd})+\frac{1}{D-2}\theta k^{c}k_{(a}\na_{|c|}k_{b)}\nn\\
&&+\frac{1}{D-1}\bar{\theta}\zeta^{c}\zeta_{(a}\na_{|c|}\zeta_{b)}
+\frac{1}{D-2}k_{a}k_{b}k^{c}\na_{c}\theta
+\frac{1}{D-1}\zeta_{a}\zeta_{b}\zeta^{c}\na_{c}\bar{\theta}.
\label{eq3null}
\eea

Taking the {\it ansatz} that the expansion $\bar{\theta}$ is constant
along the $\sigma$-direction as in the time-like case, we have
another Raychaudhuri type equation \beq
\frac{d\theta}{d\lambda}=-\frac{1}{D-2}\theta^{2}+\frac{1}{D-1}\bar{\theta}^{2}
-\sigma_{ab}\sigma^{ab}+\bar{\sigma}_{ab}\bar{\sigma}^{ab}
+\omega_{ab}\omega^{ab}-\bar{\omega}_{ab}\bar{\omega}^{ab}
-R_{ab}(k^{a}k^{b}-\zeta^{a}\zeta^{b}).\label{treqnull} \eeq
Assuming $\omega_{ab}=\bar{\omega}_{ab}$,
$\sigma_{ab}=\bar{\sigma}_{ab}$ and
a stringy strong energy condition for null case
\beq
R_{ab}(k^{a}k^{b}-\zeta^{a}\zeta^{b})\ge 0
\eeq
and exploiting the
energy-momentum tensor of the perfect fluid, we reproduce the
inequality (\ref{secstring}) in the time-like congruence of
strings. The Raychaudhuri type equation (\ref{treqnull}) for the
null strings then has a solution in the following form
\beq
\frac{1}{\theta(\tau)}\ge
\frac{1}{\theta(0)}+\frac{1}{D-2}\left(\lambda-\frac{D-2}{D-1}\int_{0}^{\lambda}
{\rm d}\lambda~\left(\frac{\bar{\theta}}{\theta}\right)^{2}\right),\label{solnnull}
\eeq where $\theta(0)$ is the initial value of $\theta$ at
$\lambda=0$. We assume again that $\theta(0)$ is negative. The
inequality (\ref{solnnull}) then implies that $\theta$ must pass
through the singularity within an affine length~\cite{hawking70}
\beq \lambda\le
\frac{D-2}{|\theta(0)|}+\frac{D-2}{D-1}\int_{0}^{\lambda} {\rm
d}\lambda~\left(\frac{\bar{\theta}}{\theta}\right)^{2}.
\label{propertimenull} \eeq

Similarly, we assume that $\omega_{ab}=\bar{\omega}_{ab}$,
$\sigma_{ab}=\bar{\sigma}_{ab}$ and $\theta\gg\bar{\theta}$ to
obtain
\beq \frac{d\omega_{ab}}{d\lambda}=-\frac{2}{D-2}\theta(\omega_{ab}
-k^{c}k_{[a}\omega_{b]c}).\label{omegataunull} \eeq
Here one notes that the twist $\omega_{bc}$ is orthogonal to the null tangent vector field
$k^{c}$ so that their inner product contraction $k^{c}\omega_{bc}$
in (\ref{omegataunull}) vanishes. We then have the above equation of the form
\beq \frac{d\omega_{ab}}{d\lambda}=-\frac{2}{D-2}\theta\omega_{ab}\label{omegatau2null} \eeq
whose solution is given by
\beq
\omega_{ab}(\lambda)=\omega_{ab}(0)\exp\left(-\frac{2}{D-2}\int_{0}^{\lambda}{\rm d}\lambda~\theta\right).
\label{omegaabnull}
\eeq
As in the case of time-like case, as the early universe evolves with the expansion rate $\theta$,
$\theta$ increases and the twist of the null stringy congruence $\omega_{ab}$ decreases exponentially.

Next, we assume that the shear $\bar{\omega}_{ab}$ is constant
along the $\sigma$-direction as in the time-like case and
$\omega_{ab}=\bar{\omega}_{ab}$,
$\sigma_{ab}=\bar{\sigma}_{ab}$ and $\theta\gg\bar{\theta}$ to
yield
\bea
\frac{d\sigma_{ab}}{d\lambda}&=&-\frac{1}{(D-2)^{2}}\theta^{2}k_{a}k_{b}
-\frac{2}{D-2}\theta h^{c}_{~(a}\sigma_{b)c}+\left(R_{c(ab)d}+\frac{1}{D-2}g_{ab}R_{cd}\right)k^{c}k^{d}
-\left(R_{c(ab)d}+\frac{1}{D-1}g_{ab}R_{cd}\right)\zeta^{c}\zeta^{d}
\nn\\
&&
+\frac{1}{(D-1)(D-2)}g_{ab}(\sigma_{cd}\sigma^{cd}-\omega_{cd}\omega^{cd})
-\frac{1}{D-2}\theta k^{c}k_{(a}\na_{|c|}k_{b)}-\frac{1}{D-2}k_{a}k_{b}k^{c}\na_{c}\theta.
\label{sigmaabdynnull}
\eea
As in the case of the time-like stringy congruence, all the shear tensor components again vanishes,
\beq
\sigma_{ab}=0,
\label{sigmaabnull}
\eeq
so that there are no shear features in the homogeneous and isotropic universe
regardless of the dynamic equation for the shear $\sigma_{ab}$ in (\ref{sigmaabdynnull}).

Finally, we consider the point particle case of the null congruence with
$\bar{B}_{ab}=\bar{\theta}=\bar{\sigma}_{ab}=\bar{\omega}_{ab}=0$
and $\omega_{ab}=0$. We assume that the strong energy condition
\beq
R_{ab}k^{a}k^{b}\ge 0\label{rkknull}
\eeq is satisfied, then we obtain
\beq
\rho+P\ge0,\label{strongpointnull}
\eeq
which is the second inequality of (\ref{strongpoint})~\cite{hawking70,wald84,carroll}. If we assume
that $\theta(0)$ is negative, the expansion $\theta(\tau)$ must go
to the negative infinity along that geodesic within a finite
affine length to yield~\cite{hawking70}
\beq
\lambda\le\frac{2}{|\theta(0)|}.
\label{lambdanullpoint}
\eeq
As for the point particle case of the twist of the null congruence, we have
$\omega_{ab}=\bar{\omega}_{ab}=\sigma_{ab}=\bar{\sigma}_{ab}=0$
so that there are no twist and shear motions in the homogeneous and isotropic universe.

\end{document}